\documentclass[%
 reprint,
 amsmath,amssymb,
 aps,
]{revtex4-2}

\usepackage{graphicx}% Include figure files
\usepackage{dcolumn}% Align table columns on decimal point
\usepackage{bm}% bold math
\usepackage{color}
\usepackage[normalem]{ulem}
\newcommand{\kst}{\vk_{\text{st}}}

\newcommand{\vk}{{\mathbf{k}}}

\newcommand{\ve}{\mathbf{e}}

\newcommand{\vE}{\mathbf{E}}
\newcommand{\vA}{\mathbf{A}}
\newcommand{\ver}{\mathbf{r}}

\newcommand{\vp}{\mathbf{p}}

\begin{document}
\preprint{APS/123-QED}
\title{Impact of the continuum Coulomb interaction in quantum-orbit-based treatments of high-order above-threshold ionization}
\author{T. Rook$^{1}$, D. Habibovi\'{c}$^{2}$, L. Cruz Rodriguez$^1$, D. B. Milo\v{s}evi\'{c}$^{2,3}$, C. Figueira de Morisson Faria$^{1}$}
\affiliation{$^1$Department of Physics and Astronomy, University College London, Gower Street, London, WC1E 6BT, UK\\$^2$
University of Sarajevo, Faculty of Science, Zmaja od Bosne 35, 71000 Sarajevo, Bosnia and Herzegovina\\$^3$Academy of Sciences and Arts of Bosnia and Herzegovina, Bistrik 7, 71000 Sarajevo, Bosnia and Herzegovina
}

\date{\today}
             
\begin{abstract}
  We perform a systematic comparison between photoelectron momentum distributions computed with the rescattered-quantum orbit strong-field approximation (RQSFA) and the Coulomb-quantum orbit strong-field approximation (CQSFA). We exclude direct, hybrid, and multiple scattered CQSFA trajectories, and focus on the contributions of trajectories that undergo a single act of rescattering.  For this orbit subset, one may establish a one-to-one correspondence between the RQSFA and CQSFA contributions for backscattered and forward-scattered trajectory pairs. We assess the influence of the Coulomb potential on the ionization and rescattering times of specific trajectory pairs, kinematic constraints determined by rescattering, and quantum interference between specific pairs of trajectories. We analyze how the Coulomb potential alters their ionization and return times, and their interference in photoelectron momentum distributions. We show that Coulomb effects are not significant for high or medium photoelectron energies and shorter orbits, while, for lower momentum ranges or longer electron excursion times in the continuum, the residual Coulomb potential is more important. We also assess the agreement of both theories for different field parameters, and show that it improves with  the  increase of the wavelength.
\end{abstract}
%\pacs{38.30.Rm,34.00} 

\maketitle

\section{\label{sec:intro}Introduction}

The laser-induced rescattering or recombination of an electron with its parent ion plays a huge role in intense-field laser-matter interaction \cite{Corkum1993,Schafer1993}. This physical picture has been instrumental in explaining a wide range of phenomena, which could not be interpreted by other means, and also in dictating the attosecond $(10^{-18}$ s) time scales for which they occurred. These processes may lead to the generation of high-order harmonics \cite{Salieres1995,Ditmire1996,Bellini1998,salieres1999}, or photoelectrons with energies well above the ionization threshold, in what is known as above-threshold ionization (ATI) \cite{Faisal1973,Agostini1979,Kruit1983}. 

Besides the myriad applications, such as attosecond light \cite{Paul2001,Hentschel2001} and electron \cite{Kim2023} pulses and photoelectron holography \cite{HuismansScience2011,Faria2020}, rescattering processes naturally call for orbit-based approaches, both classical and quantum mechanical. Quantum mechanically, electron orbits are associated with transition amplitudes, and describe the several pathways an electron can take up to the detector. This is the key ingredient in several semi-analytic approaches, which have been formulated to describe the dynamics of the liberated electron under the influence of the driving field and the Coulomb potential of the residual ion \cite{Lewenstein1994,Becker1997}.  If the process is analyzed in terms of the Feynman path-integral formalism \cite{Salieres2001}, the ionization amplitude is written as the sum of the contributions of different quantum orbits \cite{Kopold2000,Milosevic2006}. These quantum orbits have a well-known spacetime evolution, so the method allows us to investigate different options available to the freed electron. 

The most traditional and widespread of such approaches is the strong-field approximation (SFA). The SFA assumes that the driving field is so strong that the influence of the parent ion on the liberated electron can be neglected during the electron propagation upon eventual rescattering \cite{Milosevic2003,MilosReviewATI,Krausz2009,Becker2012,Symphony}. This approximates the continuum by field-dressed plane waves \cite{Keldysh1965,Faisal1973,Reiss1980}, which are analytically tractable, and rescattering is incorporated by constructing a Born-type series around these solutions \cite{Lewenstein1995,Lohr1997}.  This leads to a clear-cut definition of scattering, which has influenced the main way of thinking in the research area for decades.

Specifically for ATI, besides the scenario in which the liberated electron goes directly to the detector (the so-called direct electrons), the oscillatory character of the applied field can return the electron in the vicinity of its parent ion and the rescattering may occur. If rescattering happens, the process is denoted as high-order ATI (HATI) and its signature is a long plateau with comparable peak intensities \cite{Paulus1994} (for review see, e.g., \cite{Becker2002Review,Agostini2012,Becker2018}). These electrons are known as rescattered electrons.  Quantum-orbit methods were employed within the SFA to investigate the direct electrons (see for example Ref.~\cite{MilosReviewATI}), while the solutions for the rescattered electrons were systematically presented and classified in Ref.~\cite{Milosevic2014} for linearly polarized and in Ref.~\cite{Milosevic2016} for bicircular driving fields.

The SFA theory was applied to the HATI process in Refs.~\cite{Yang1993,Lewenstein1995}, but the role of the residual Coulomb potential remained an open question. At first, the influence of the Coulomb potential was accounted for through the Born series analyzing the influence of the screening parameter \cite{Milosevic1997}, while later approaches incorporate Coulomb distortions in the resulting scattering waves \cite{Arbo2008}, combined with the Born series \cite{Milosevic1998}.  Eventually, the continuum propagation was modified  \cite{Popruzhenko2008a,Popruzhenko2008d,Smirnova2008,Torlina2013}, although not all groups of orbits leading to substantial contributions were taken into consideration \cite{Popruzhenko2008a,Popruzhenko2008d}, or approximations have been made on the scattering angle \cite{Smirnova2006,Smirnova2008,Torlina2013}. For the main methods beyond the SFA see the review article \cite{Faria2020}.

A Coulomb-distorted continuum invites a wide range of questions, such as whether a distinction between ``direct" and ``rescattered" electrons can still be made and, if so, what would be its validity range. Many features identified in photoelectron holography, such as the fan-shaped structure that occurs near the ionization threshold \cite{Rudenko2004,Maharjan2006}, or the spider-like structure near the polarization axis \cite{HuismansScience2011}, were observed in experiments and in the full numerical solution of the time-dependent Schr\"odinger equation, but not in the SFA. More recent examples are spiral-like structures \cite{Maxwell2020}, or twisted holographic patterns that have been predicted for elliptically polarized fields if the Coulomb potential is included \cite{Kim2022}. 

An important breakthrough occurred in the past decade, with the development of strong-field path-integral approaches that incorporate the driving field and the Coulomb potential on equal footing \cite{Yan2010,Yan2012,Li2014,Lai2015a,Shilovski2016,Shilovski2018}. These approaches do not resort to Born-type expansions, which means that the distinction between ``direct" and ``rescattered" electrons is blurred. This is expected as, in reality, the interplay between the residual binding potential and the driving field is highly non-trivial, which invites many questions. Would an electron, being deflected by the potential, but whose perihelion is larger than the Bohr radius be direct or rescattered? For instance, in \cite{Lai2017,Maxwell2017}  it was shown that a fan-shaped holographic structure could only be created if the Coulomb tail was accounted for, but for all purposes, the interfering orbits were viewed as ``direct". Furthermore, in \cite{Yan2010}, whole sets of orbits have been encountered which had no SFA counterpart. What about the scenarios in which an electron is trapped or recaptured by the ionic potential \cite{Popruzhenko2017,Zhao2019,Cao2021}? Above all, how to compare theories whose structure is so different?

In our earlier publications \cite{Lai2015a,Lai2017,Maxwell2017,Maxwell2018,Kim2022} the comparison between the results obtained using the SFA and the Coulomb quantum-orbit strong-field approximation (CQSFA) was made without considering the structural differences between the theories.
For example, in the SFA theory, the intermediate electron momentum changes abruptly at the moment of rescattering which happens when the electron is at the position of the core. In the CQSFA, this is not the case, i.e., the trajectory can go very close to zero, but will never be zero, while the change of the photoelectron momentum is continuous. Furthermore, we focused on what the SFA leaves out, instead of looking for similarities. The strong-field approximation was presented as a limit for the CQSFA orbits in specific parameter ranges \cite{Maxwell2018}, or comparisons with direct SFA electrons were performed to emphasize the deflection caused by the residual potential \cite{Maxwell2017,Kim2022}. Up to recently, the CQSFA was solved as a purely boundary problem, which required some pre-knowledge of the orbits' dynamics. This excluded entire sets of orbits, which were revealed in a recent publication in which hybrid and forward implementations of the CQSFA were made \cite{rodriguez2023}. These orbits led to annular intra-cycle fringes that resembled those found for rescattered SFA orbits \cite{Spanner2004}.

In the present paper, we systematically compare the results obtained using our theories. We discuss how the CQSFA  trajectories which mimic the behavior of the SFA trajectories can be isolated, and how good is the agreement between the partial contributions to the probability density. Also, we discuss how these differences depend on the photoelectron energy by tracking the phase accumulated due to the Coulomb potential. Our focus will be on the rescattered regime, as a comparison with direct SFA electrons has been performed elsewhere \cite{Lai2017,Maxwell2017}. 

To keep the focus on how far a Born-type approach goes in reproducing the full continuum dynamics, we have opted for the simplest possible system: hydrogen in a linearly polarized monochromatic field.  Simulating a real experiment may require focal averaging, depletion, bound-state dynamics, and a realistic pulse.  These effects are potential sources of incoherence \cite{Maxwell2021a}, and they may skew the interference patterns one is trying to assess (see \cite{Maxwell2015,Maxwell2016} for examples in a two-electron scenario), and modify the dynamical constraints associated with rescattering. Assessing all these additional effects would detract from the main purpose of the present work. However, it is known that for pulses longer than 10 optical cycles a flat envelope is a good approximation  \cite{Milos2006}. Furthermore, the CQSFA has been successfully employed to explain experiments in rare gases such as xenon \cite{Maxwell2020} and argon \cite{Werby2021,Werby2022}, as well as to detect parity in diatomic molecules \cite{Kang2020}. In the modeling of these experiments, a monochromatic wave was assumed, and, still, the theory reproduced subtle effects such as multi-path holographic interference \cite{Werby2021,Werby2022}. 

The paper is organized as follows. In Section~\ref{sec:backgd} we present the two versions of the saddle-point method: one based on the SFA and one based on the Coulomb-corrected SFA. In Sec.~\ref{sec:mapping} we discuss how the trajectories obtained in the SFA theory can be extracted from the CQSFA theory. Moreover, we discuss the behavior of some of the trajectories which do not have an SFA analog. In addition, we briefly present the classification of the quantum orbits for both theories. In Section~\ref{sec:rescatter} we present our numerical results and discuss the similarities and differences between the two theories. Finally, Section~\ref{sec:conclusion} contains our main conclusions. The atomic units are used unless otherwise stated.

\section{\label{sec:backgd}Background}

The probability amplitude for the transition from the initial bound state $|\psi_0\rangle$ to the free final state $|\psi_{\vp_f}\rangle$ with momentum $\vp_f$ is given by \cite{Becker2002Review}
\begin{equation} \label{probamp2}
M_{\vp_f}=-i\lim_{t\rightarrow\infty}\int_{-\infty}^t dt_0\langle \psi_{\vp_f}(t)|U(t,t_0)\ver\cdot\vE(t_0)|\psi_0(t_0)\rangle,
\end{equation}
where $U(t,t_0)$ is the evolution operator which corresponds to the total Hamiltonian $H(t)=H_0+H_I(t)$ of the system. Here $H_0$ is the time-independent part which describes the electron exposed to the atomic potential, while the time-dependent part $H_I(t)=\ver\cdot\vE(t)$ corresponds to the interaction of the electron with the applied field $\vE(t)$.

\subsection{Strong Field Approximation}
The evolution operator $U(t,t_0)$ can also be written as
\begin{equation}\label{Usfa}
U(t,t_0)=U_{\mathrm{F}}(t,t_0)-i\int_{t_0}^t dt'' U(t,t'')V(\ver)U_{\mathrm{F}}(t'',t_0),
\end{equation} 
where $U_{\mathrm{F}}(t,t_0)$ is the Volkov evolution operator and $V(\ver)$ is the atomic potential. Using only the first term of the previous expansion we obtain the amplitude of the direct electrons
\begin{equation} \label{probampdir}
M_{\vp_f}^{(0)}=-i\int_{-\infty}^{\infty} dt_0\langle \vp_f+\vA(t_0)|\ver\cdot\vE(t_0)|\psi_0\rangle e^{iS_d(\vp_f;t_0)},
\end{equation}
with  $\vA(t)$ being the vector potential of the field $\vE(t)=-d\vA(t)/dt$ and $S_d(\vp_f;t_0)=S_{\vp_f}(t_0)+I_pt_0$ being the modified action for the direct electrons with $S_{\vp_f}(t)=\int^t dt'[\vp_f+\vA(t')]^2/2$ and $I_p$ the ionization potential. The free final state $|\psi_{\vp_f}\rangle$ is approximated by a plane wave. Finally, $t_0$ is the ionization time.
On the other hand, the second term from Eq.~\eqref{Usfa} gives the amplitude of the rescattered electrons
\begin{eqnarray} \label{probampres}
M_{\vp_f}^{(1)}=-\int_{-\infty}^{\infty} dt_0\int_{t_0}^{\infty}dt\int d\vk\langle \vp_f|V(\ver)|\vk\rangle\nonumber \\
\times\langle \vk+\vA(t_0)|\ver\cdot\vE(t_0)|\psi_0\rangle e^{iS_r(\vp_f,\vk;t,t_0)}.
\end{eqnarray}
Here  $S_r(\vp_f,\vk;t,t_0)=S_{\vp_f}(t)-S_{\vk}(t)+S_{\vk}(t_0)+I_pt_0$ is the modified action for the rescattered electrons, while $t$ is the rescattering time. For the remainder of this paper, this contribution shall be denoted by SFA.

The integral which appears in Eq.~\eqref{probampres} can be solved numerically or using the saddle-point (SP) method. The stationary condition which corresponds to the momentum $\vk$ is satisfied for $\kst(t_0,t)=-\int_{t_0}^t \vA(t')dt'/(t-t_0)$, while the conditions which correspond to the ionization $\partial S_{r}(\vp_f,\kst;t,t_0)/\partial t_0=0$ and rescattering time $\partial S_{r}(\vp_f,\kst;t,t_0)/\partial t=0$ lead to the following SP equations
\begin{equation}\label{sphati1}
[\kst+\vA(t_0)]^2=-2I_p,
\end{equation}
\begin{equation}\label{sphati2}
[\kst+\vA(t)]^2=[\vp_f+\vA(t)]^2,
\end{equation}
which represent the energy-conservation conditions at the ionization and rescattering times, respectively. The solutions were classified previously in \cite{milosevic2002} for the high-order harmonic generation (see Sec.~\ref{sec:mapping}). The modified SP method (see Appendix B in Ref.~\cite{MilosReviewATI}) leads to the following expression for the rescattering amplitude 
\begin{eqnarray}\label{Thatisp}
M_{\vp_f}^{(1),\mathrm{SP}}&=&2i\pi^{3/2}\kappa^{5/2}\sum_{\{t_{0s},t_s\}} \frac{\langle \vp_f|V|\kst\rangle}{[i(t_s-t_{0s})]^{3/2}} \nonumber \\
&& \times \frac{1}{S''_{rs0}} \left(\frac{2\pi i}{S''_{rs}}\right)^{1/2}e^{iS_{rs}},
\end{eqnarray}
where we introduced $S_{rs}=S_r(\vp_f,\kst;t_s,t_{0s})$, $S''_{rs0}=\partial^2S_{r}/\partial t_0^2|_{t_{0s},t_s}$ and $S''_{rs}=\partial^2S_{r}/\partial t^2|_{t_{0s},t_s}$, $\kappa=(2I_p)^{1/2}$, and the sum is over the solutions $t_{0s}$ and $t_s$ of the SP equations \eqref{sphati1} and \eqref{sphati2}.

\subsection{Coulomb Quantum-Orbit Strong Field Approximation}
For the approach which takes into account the Coulomb potential, the probability amplitude \eqref{probamp2} can be written as 
\begin{eqnarray} \label{probamp3}
M_{\vp_f}&=&-i\lim_{t\rightarrow\infty}\int_{-\infty}^t dt_0\int d\widetilde{\vp}_0 \langle \widetilde{\vp}_f(t)|U(t,t_0)|\widetilde{\vp}_0\rangle \nonumber \\&& \times \langle\widetilde{\vp}_0|\ver\cdot\vE(t_0)|\psi_0(t_0)\rangle,
\end{eqnarray}
with $|\widetilde{\vp}_f(t)\rangle=|\psi_{\vp_f}(t)\rangle$, and $\widetilde{\vp}_0=\vp_0+\vA(t_0)$ and $\widetilde{\vp}_f(t)=\vp_f+\vA(t)$ being the initial and final velocity of the electron at the times $t_0$ and $t$, respectively. The matrix element $\langle \widetilde{\vp}_f(t)|U(t,t_0)|\widetilde{\vp}_0\rangle$ can be calculated using the path integral method, leading to the following expression \cite{Kleinert2009,Milosevic2013JMP,Lai2015a,Milosevic2017}
\begin{eqnarray} \label{probamp4}
M_{\vp_f}&=&-i\lim_{t\rightarrow\infty}\int_{-\infty}^t dt_0\int d\widetilde{\vp}_0\int_{\widetilde{\vp}_0}^{\widetilde{\vp}_f(t)}\mathcal{D}'\widetilde{\vp} \int\frac{\mathcal{D}\ver}{(2\pi)^3} \nonumber \\&& \times e^{iS(\widetilde{\vp},\ver,t,t_0)} \langle\widetilde{\vp}_0|\ver\cdot\vE(t_0)|\psi_0\rangle,
\end{eqnarray}
where $\mathcal{D}'\vp$ and $\mathcal{D}\ver$ are the integration measures for the path integrals, the prime indicates a restriction, and the action reads
\begin{equation}\label{cqsfaAction}
    S(\widetilde{\vp},\ver,t,t_0)=I_pt_0-\int_{t_0}^t[\dot \vp(t')\cdot \ver(t')+H(t')]dt',
\end{equation}
with $H(t')=H[\ver(t'),\vp(t'),t']=[\vp(t')+\vA(t')]^2/2+V[\ver(t')]$. The stationary conditions for the variables $t_0$, $\vp$ and $\ver$ lead to the equation
\begin{equation}\label{cqsfa1}
    [\vp(t_0)+\vA(t_0)]^2+2V[\ver(t_0)]=-2I_p,
\end{equation}
as well as to the classical equations of motion of the electron
\begin{equation}\label{cqsfa23}
    \dot\vp=-\nabla_r V[\ver(t')],\quad \dot \ver=\vp+\vA(t').
\end{equation}
The integral and the saddle-point equations are solved using a two-pronged contour. The first part of the contour is chosen along the imaginary-time axis, from $t'$ to $\mathrm{Re}\text{ }t'$, and the second part of the contour is taken along the real-time axis from $\mathrm{Re}\text{ }t'$ to a final time $t\rightarrow \infty$. In the first part of the contour, the momentum is assumed to be constant and the binding potential in Eq.~(\ref{cqsfa1}) is neglected. This makes the sub-barrier contribution to the action dependent on the tunnel trajectory.
\begin{equation}\label{eq:rTun}
	\textbf{r}_0(\tau)=\displaystyle\int_{t'}^{\tau}[\textbf{p}_0+\textbf{A}(\tau')]d\tau',
\end{equation}
which can be used to define the tunnel exit 
\begin{equation}
z_0 = \mathrm{Re}[r_{0||}(t'_r)],
\label{eq:tunnelExitGeneral}
\end{equation}
where $r_{0||}(t'_r)$ is the component of the tunnel trajectory [Eq.~\eqref{eq:rTun} for $\tau=t'_r$] parallel to the driving-field polarization. One should note that setting the tunnel exit to be real is an approximation, which renders the continuum propagation equations real. This approximation leads to practically no changes in the interference patterns and has been used in almost all our publications. Solving the full complex problem will involve branch cuts and has been discussed elsewhere \cite{Maxwell2018b}. 

A classification of the solutions to the system of equations \eqref{cqsfa1} and \eqref{cqsfa23} was introduced in \cite{Lai2015a} based upon the initial conditions and the final momentum of the photoelectron trajectory. However, this classification was found to be incomplete \cite{rodriguez2023} because more information about the intermediate momentum and position is necessary for a full characterization of the qualitative nature of the trajectory (see Sec.~\ref{sec:mapping}). Using the solutions $t_{0s}$, $\vp_s$, and $\ver_s$ of these Coulomb-corrected SP equations, the probability amplitude can be written as 
\begin{eqnarray}
M_{\vp_f}^{\text{CSP}}&\propto&\lim_{t\rightarrow\infty}\sum_s D^{-1/2}\mathcal{C}(t_{0s})e^{iS_s-i\pi\nu_s/2},
\end{eqnarray}
where $S_s=S(\widetilde{\vp}_s,\ver_s,t,t_{0s})$, $D=\text{det}\left[\partial\vp_s(t)/\partial \vp_s(t_{0s})\right]$,
\begin{equation}
    \mathcal{C}(t_{0s})=\sqrt{\frac{2\pi i}{S''_0}}\langle \vp_f+\vA(t_{0s})|\ver\cdot \vE(t_{0s})|\psi_0\rangle,
\end{equation}
$S''_0=\partial^2S(\widetilde{\vp}_s,\ver_s,t,t_{0s})/\partial t_{0s}^2$, and $\nu_s$ is the Maslov phase associated with a solution $s$ as calculated by the prescription given in \cite{carlsen_advanced_2023} and \cite{brennecke2020gouy}.
Here, we use the hybrid CQSFA implementation discussed in \cite{rodriguez2023}.

In the following section, we compare our results obtained  using the SFA and CQSFA theories using the example of the hydrogen atom exposed to a linearly polarized field. The ionization potential is $0.5$ a.u. Our driving field is $\vE(t)=E_0\sin(\omega t)\hat{\ve}_x$ with $E_0$ being the amplitude, and we define the emission angle $\theta_e$ as the angle between the final photoelectron momentum and the unit vector $\hat{\ve}_x$.

\section{Momentum Mapping and Orbit Correspondence}\label{sec:mapping}

In this section, we explain how the two theoretical models are related to each other. The SFA and CQSFA theories are different in a way that the SFA  is a Born-expansion-based theory, meaning that the electrons which go directly to the detector, as well as the electrons which experience one or more rescattering events, are clearly 
identifiable. The continuum is approximated by laser-dressed plane waves and the electron rescatters with the potential when it returns to the site of its release, namely the origin. Spatially, this assumption locates the interaction at a single point.

On the other hand, the CQSFA approach treats the Coulomb and the laser-field potentials equally. Because of the spatial range of the Coulomb potential, it is not straightforward to determine whether the electron can be viewed as direct, undergoing a mere deflection, a soft collision, or a hard collision. Furthermore, the potential may trap the outgoing electron, or lead to multiple collisions. This makes the difference between the direct and rescattered electrons blurred, and also means that not every orbit in the CQSFA will exhibit an SFA counterpart. Therefore, we must identify those that do.  The main goal of this section is to understand what types of orbits can be compared between these methods and the procedure one must follow in order to do this. 

\subsection{Orbit Classification}
\label{sec:orbitclass}
We start with a short reminder about the classification of the SP solutions for both theories. In the SFA theory, the rescattering solutions occur in pairs associated with a longer and shorter orbit, which coalesce for maximal classically allowed rescattering energies. These energies are associated with sharp decreases in the photoelectron signal, which are known as `cutoffs' \cite{Paulus1994JPhysB}. They are classified into two groups: the backward- and forward-scattering solutions. In the former case, the electron is rescattered in approximately the opposite direction with respect to the direction of its motion before the scattering event, while in the latter case, the electron is only deflected during this event. Therefore, within the SFA, the parameter used to enforce this distinction is the rescattering angle.

The backward-scattering solutions are classified using the multiindex $(\alpha,\beta,m)$. For the return times within one optical cycle, there are infinitely many solutions with different travel times $\tau=t-t_0$. The index $m$ gives the approximate value of the travel time in units of laser-field period $T=2\pi/\omega$, $\omega$ being the angular frequency. For $-(m+1/2)T<t_0<-(m-1/2)T$, $m=0,1,2,\ldots$, there are two pairs of solutions distinguished by the index $\beta$ in such a way that $\beta=-1$ ($\beta=+1$) corresponds to the longer (shorter) travel time. Furthermore,  the two solutions of one pair are distinguished by the index $\alpha$. The long (short) solution is denoted by $\alpha=-1$ ($\alpha=+1$). 

The forward-scattering solutions were classified in \cite{Milosevic2014} using the  multiindex $(\mu,\nu)$ in such a way that $\mu=0,1,2,\ldots$, measures the travel time, while index $\nu=\pm 1$ is used to distinguish the two solutions of one pair. The positions  of the cutoff for these  solutions were classically found in \cite{Becker2014a}, while those positions with quantum  corrections were given in \cite{milosevic2016a}. Besides the shortest and the second-shortest pair, other forward-scattering solutions contribute only in a low-energy part of the spectra [$E_{\vp_f}<0.04U_p$, $U_p=E_0^2/(4 \omega^2)$ being the ponderomotive energy]. 

For the CQSFA theory, the classification of the solutions was introduced in \cite{Lai2015a,Maxwell2017} based upon the initial conditions and the final momentum of the photoelectron trajectory. The solutions with $p_{f,y}p_{0,y}>0$ are denoted as `orbit 1' for $z_0p_{f,x}>0$ and `orbit 2' for $z_0p_{f,x}<0$, while for $p_{f,y}p_{0,y}<0$ the solutions are denoted as `orbit 3' for $z_0p_{f,x}<0$ and `orbit 4' for $z_0p_{f,x}>0$, where $z_0$ is the tunnel exit. 
In those early publications, orbit 1 was identified with a direct SFA orbit in which the electron is freed in the direction of the detector, orbits 2 and 3 behave as field-dressed hyperbolas starting half a cycle earlier or later, with the difference that orbit 2 interacts less with the core and can still be viewed as ``direct", while orbit 3 has a hybrid character. Orbit 4 goes around the core before reaching the detector, and is identified as backscattered. In more recent publications \cite{Maxwell2018,rodriguez2023}, however, we found that these distinctions are blurred and a wide range of behaviors for each of the four types of orbit can be identified by applying spatial filters. If the orbit's distance of closest approach $r_c$ is larger than the radius $r_T$ determined by the tunnel exit, the CQSFA orbit is considered `direct'. If the orbit's perihelion lies between the tunnel exit and the Bohr radius, that is, $r_T \ge r_c \ge r_0$, we refer to the orbit as `soft scattered', while if $r_c < r_0$ the CQSFA orbit is `hard scattered'. A thorough analysis using these filters has been performed in \cite{rodriguez2023}. 

Here, instead of using a spatial filter, in order to compare the CQSFA with the SFA, we eliminate all orbits which do not have analogous SFA solutions by ruling out multiple scattering events or single scattering events induced significantly by the long range of the Coulomb potential. The simple criterion which was used is keeping only the solutions which correspond to the trajectories that cross the $r_y=0$ axis once. It turns out that this criterion allows one to recover families of solutions which may be compared with the single-rescattering SFA solutions. Additional constraints have been imposed by matching the CQSFA rescattering times to their SFA counterparts, as will be discussed below. 

Following the approximations stated above, the rescattering solutions found in the SFA theory can be reproduced from the orbit 3 and orbit 4 solutions from the CQSFA theory. Orbit 1 and 2 solutions mainly correspond to the direct electrons. 
The relation between the solutions obtained using our theories is summarized in Table~\ref{tab:orbits}. 
\begin{table}[!htbp]
    \centering
   \begin{tabular}{||c| c| c| c||} 
 \hline
 Orbit & $z_0p_{f,x}$ & $p_{f,y}p_{0,y}$ & SFA orbit \\ [0.5ex] 
 \hline\hline
 1 & $+$ & $+$ & direct \\ 
 \hline
 2 & $-$ & $+$ & direct \\
 \hline
 3 & $-$ & $-$ & $(\beta,m)=(1,1),(1,2)$, \\
 &&&forward-scattering \\
 \hline
 4 & $+$ & $-$ & $(\beta,m)=(-1,0),(-1,1)$ \\ [1ex] 
 \hline
\end{tabular}
    \caption{Main types of orbits identified using CQSFA approach. The $+$ ($-$) sign on each cell indicates a positive (negative) value of the product. The fourth column contains the SFA orbits which can be extracted from the corresponding CQSFA orbits.}
    \label{tab:orbits}
\end{table}
\subsection{Extracting SFA solutions}
Here, the SFA orbits that can be extracted from the corresponding CQSFA orbits are given in the fourth column. For the SFA solutions $(\beta,m)=(1,0)$ \cite{Milosevic2014} we have not found the CQSFA analogue. Finally, it is worth  mentioning that the CQSFA orbits are classified using the boundary conditions of the trajectories that specify the four types of orbits, while the SFA orbits are classified using the value of the electron's velocity just before rescattering. 

\begin{figure}[!htbp]
    \centering
    \includegraphics[width=0.99\linewidth]{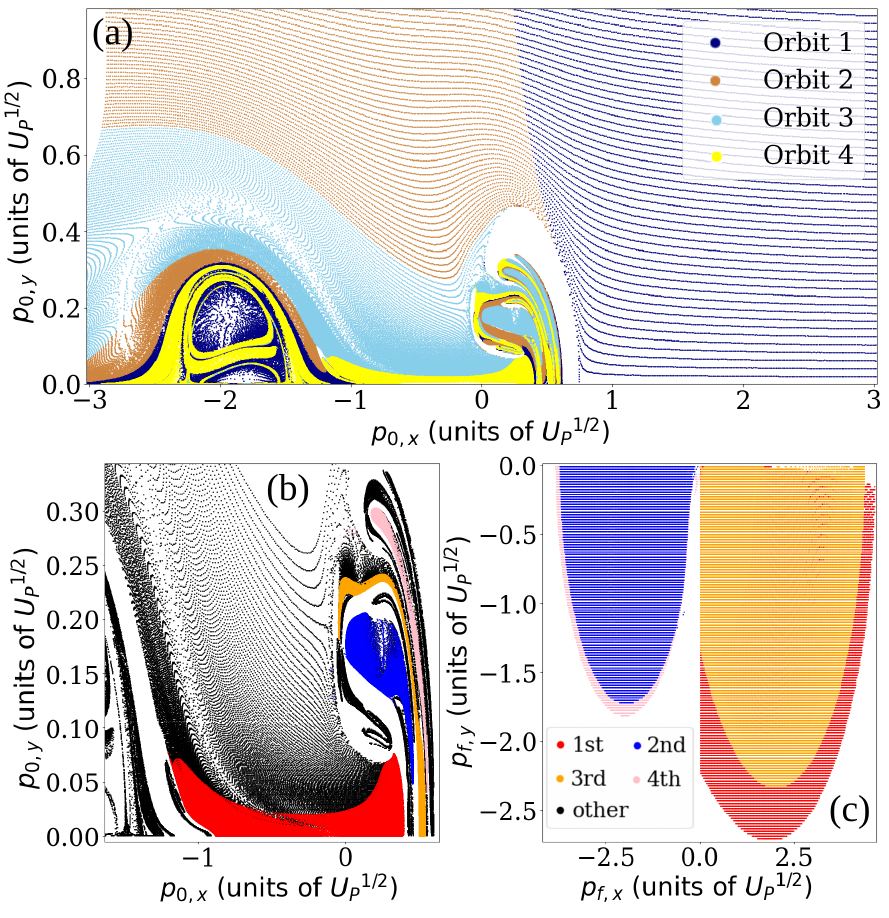}
    \caption{Panel (a): The values of the initial momentum of semiclassical trajectories calculated using the CQSFA whose final momenta belong to a uniform grid. Different types of orbits are presented with different colors as indicated in the legend. The results are obtained for the second-half cycle of the monochromatic field. Panels (b) and (c): The  values  of  the initial [panel (b)] and final [panel (c)] momentum of semiclassical trajectories for orbits 3 and 4, obtained  using the CQSFA theory, which have analogous counterparts in the SFA backward-scattered solutions. They are coded in different colors as indicated in the legend in panel (c). The intensity of the applied field is $I=E_0^2=2\times 10^{14}$~W/cm$^2$, while the angular frequency $\omega$ corresponds to the wavelength of $800$~nm.}
    \label{fig:momMap}
\end{figure}

Now we illustrate how we extracted the earlier-mentioned SFA solutions from the CQSFA theory. In Fig.~\ref{fig:momMap}(a) we  present the  values  of  the initial momentum of semiclassical  trajectories calculated using the CQSFA theory whose final momenta belong to a uniform grid. Different types of orbits (see Table~\ref{tab:orbits}) are coded with different colors as indicated  in the  legend. For simplicity, we have considered only solutions with the  ionization time in the second half cycle. The half-cycle symmetry that is inherent to this field \cite{Rook2022,Habibovic2021} guarantees that there will be no loss of information. Both half cycles have been considered in \cite{rodriguez2023}. The solutions with the final momentum in the first (second) quadrant of the final momentum plane correspond to orbit 1 (2). On the other hand, the solutions with the final momentum in the third (fourth) quadrant of the final momentum plane, shown in Fig.~\ref{fig:momMap}, correspond to orbit 3 (4). The island close to the origin leads to the rescattering ridges, while the island centered at $(p_{0,x},p_{0,y})=(-2U_p^{1/2},0.2U_p^{1/2})$ gives rise to caustics \cite{rodriguez2023}. Both of these islands are visible in Fig.~\ref{fig:momMap}(a), while in Fig.~\ref{fig:momMap}(b) only the central island, which is responsible for the ridges shown in Fig.~\ref{fig:momMap}(c), has been presented.

Using specific subsets of the initial momentum points, rescattered trajectories analogous to those obtained in the SFA can be obtained. In Fig.~\ref{fig:momMap}(b) we display these subsets in such a  way that the red (gray), blue (dark gray), orange (light gray) and pink (bright gray) regions lead to the backward-scattering SFA solutions listed in order of the increasing travel time. For example, the solutions represented by the red (gray) points correspond to the backward-scattered SFA solutions with the shortest travel time. Since these solutions can all be classified as orbit 3 or orbit 4, only the solutions classified as such are displayed in Fig.\ref{fig:momMap}(b) and Fig.\ref{fig:momMap}(c). For the CQSFA we choose the rescattering time to be the time at which the trajectory reaches its perihelion. Physically, this is the most appropriate assumption one can make to compare with the SFA, which is a Born-type expansion for which rescattering times are well defined according to Eq.~\eqref{sphati2}. 

The range of these solutions regarding the final momentum is shown in Fig.~\ref{fig:momMap}(c), with rescattering ridges marking the maximal classical momenta associated with specific orbits. The rescattering ridges formed by the first- and third-shortest solutions intersect the $p_{f,x}=0$ axis, while this is not the case for the second- and fourth-shortest solutions. The first- and third-shortest solutions with the final momentum $p_{f,x}<0$ (not shown in the figure) correspond to the SFA forward-scattering solutions. Similar solutions do not exist for the second- and fourth-shortest pairs so that the forward-scattering SFA solutions can only be extracted from the orbit 3 CQSFA solutions. The solutions represented by the black dots [see Fig.~\ref{fig:momMap}(b)] account for the remaining CQSFA trajectories. These include the forward-scattering solutions and the solutions which include multiple scattering events.  
A detailed study of the initial-to-final momentum mapping of the CQSFA orbits, with and without SFA counterparts, has been performed in \cite{rodriguez2023}, using different versions of the CQSFA and the spatial filtering mentioned in Sec.~\ref{sec:orbitclass}. 

\subsection{Counterexamples: orbits with no SFA analogy}

Despite the fact that the direct SFA orbits can be extracted from orbits 1 and 2, this does not mean that orbits classified as 1 and 2 correspond only to the direct electrons. To illustrate this, we show a few examples of the rescattering trajectories corresponding to either orbit 1 or orbit 2 CQSFA solutions.  
\begin{figure}[!htbp]
    \centering
    \includegraphics[width=0.7\linewidth]{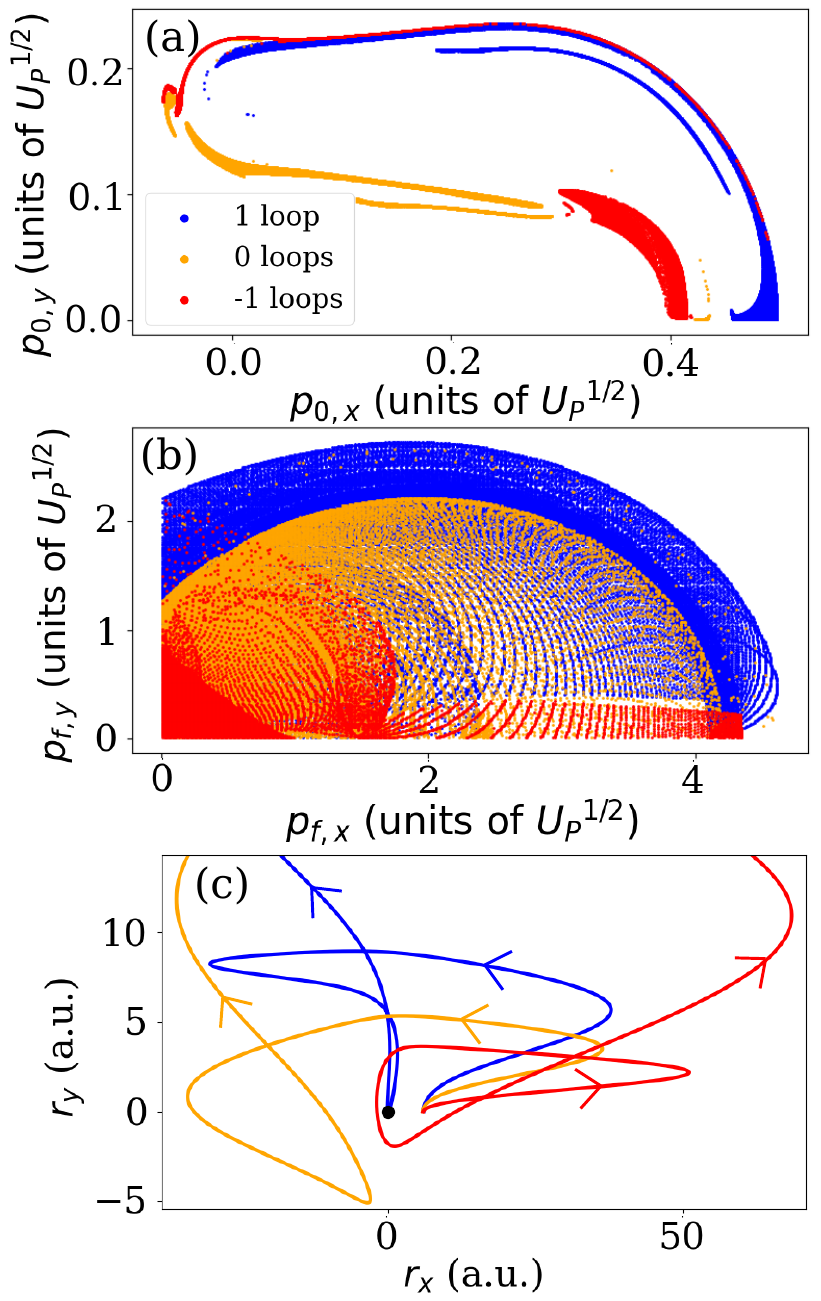}
    \caption{The values of the initial [panel (a)] and final [panel (b)] photoelectron momentum of the semiclassical orbit 1 trajectory calculated using the CQSFA theory for the situation in which the trajectories make one loop in the clockwise [blue (dark gray) regions] or counterclockwise [red (gray) regions] direction around the core, as well as for the trajectories which do not make any loop around the core [orange (light gray) regions]. Panel (c): The examples of orbit 1 trajectories for given values of the initial conditions.}
    \label{fig:momMap1}
\end{figure}
In Fig.~\ref{fig:momMap1} we present the values of the initial [panel (a)] and final [panel (b)] photoelectron momentum of the semiclassical orbit 1 trajectories calculated using the CQSFA theory. Among all solutions, we extract those  which correspond to the trajectories which make a loop in the  clockwise [blue (dark gray) regions] or counterclockwise [red (gray) regions] direction around the core. The regions which  correspond to the trajectories which do not make a loop around the core are denoted  by orange (light gray) color. The three regions of the final momentum which correspond to these trajectories partiality overlap [see Fig.~\ref{fig:momMap1}(b)]. However, the ridges of these trajectories can still be easily noticed. The ridge associated to the trajectory which makes a loop in the  clockwise direction has the highest energy, while the lowest-energy ridge corresponds to the trajectory which makes a loop in the counterclockwise direction around the core. The examples of these trajectories are shown in Fig.~\ref{fig:momMap1}(c). The values of the initial-momentum components for the  trajectory which makes a loop in the  clockwise (counterclockwise) direction are $p_{0,x}=0.00214$ and $p_{0,y}=0.212$ ($p_{0,x}=0.3679$ and $p_{0,y}=0.0711$), while for the trajectory which does not make a loop around the core, these values are $p_{0,x}=-0.0254$ and $p_{0,y}=0.135$, in units of $U_p^{1/2}$. For clarity, although the field dressing means that initially some of the trajectories shown in Fig.~\ref{fig:momMap1}(c) travel into the 2nd quadrant of the plane, they are indeed orbit 1 trajectories and will eventually move into the 1st quadrant outside of the position ranges shown in the figure. Clearly, these are rescattering trajectories even though they correspond to the orbit 1 solutions. Additionally, the radial vector of the perihelion of the trajectory which loops around the core in a counter-clockwise direction [see the red (gray) line in Fig.~\ref{fig:momMap1}(c)] makes an angle of $55^\circ$ with the negative $r_x$ axis. This shift of the perihelion is what allows the photoelectron to scatter into the first quadrant and is an effect which relies entirely upon the presence of a long-range Coulomb potential. Similar conclusions are valid for the orbit 2 solutions.  

The examples discussed above focus on the specific scenarios for which there is hard scattering, in the sense that the orbit's distance of closest approach is smaller than that defined by the tunnel exit. However, in previous publications we have provided examples of many orbits which do not have SFA counterparts. For instance, in  \cite{Lai2017} we have shown that the near-threshold fan-shaped structure encountered in experiments occurs due to a Coulomb distortion of orbits 2, which is angle-dependent and resembles a laser-dressed Kepler hyperbola. Upon interference with orbits 1, it gives rise to a radial set of fringes centered at $(p_{f,x},p_{f,y})=(0,0)$. A direct orbit 2 in the SFA sense, instead of the CQSFA orbit, leads to nearly vertical structures,  and fails to reproduce the fan \cite{Maxwell2017}. Furthermore, orbit 3 cannot always be understood as ``forward scattered", and under some circumstances exhibits a hybrid character. This is important for an accurate description of the spider-like fringes \cite{Maxwell2017,Maxwell2018}. In \cite{Maxwell2018}, we reported several overlooked holographic structures, whose existence is determined by the existence of the residual potential in the continuum. One of these structures, the spiral, has been identified in experiments \cite{Maxwell2020}, and has been shown not to be reproduced by either direct or rescattered ATI using the standard SFA.

\begin{figure*}[!htbp]
    \hspace{-7 cm}$(\beta,m)=$ $(1,2)$ $~~~~~(-1,1)$ $~~~~~(1,1)$ $~~~~~(-1,0)$

    \centering
    \includegraphics[width=0.7\linewidth]{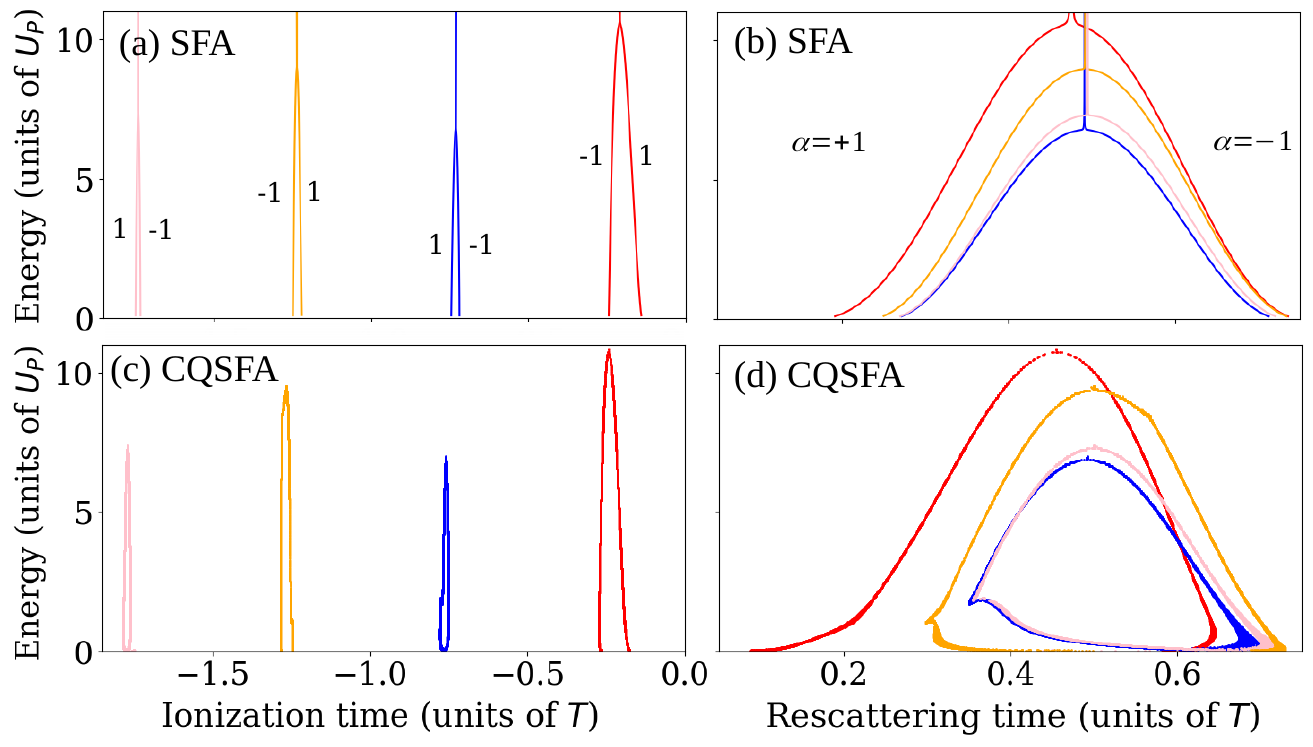}
    \caption{Ionization [panels (a) and (c)] and rescattering [panels (b) and (d)] times for the SP solutions  relevant in the medium- and high-energy regions of the spectrum calculated using the SFA [panels (a) and (b)] and  CQSFA [panels (c) and (d)] theories. The intensity and frequency of the applied field are the same as in Fig.~\ref{fig:momMap}. The emission angle is $\theta_e=0^\circ$. The values of the multiindex $(\beta, m)$ are indicated above panel (a), while the values of index $\alpha$ are indicated within panels (a) and (b).}
    \label{fig:backScatterTimes}
\end{figure*}

\section{Rescattered electrons - SFA vs CQSFA\label{sec:rescatter}}
In this section, we present numerical results for the ionization and rescattering times as well as the photoelectron momentum distributions obtained using the strong-field approximation and the Coulomb quantum-orbit strong-field approximation theories.

\subsection{Times and orbits}
First, we investigate and classify the solutions for the ionization and rescattering times obtained using the SFA and CQSFA theories. The low-energy region is dominated by the contribution of direct electrons which means that the rescattered solutions, which we consider here, are most relevant in the medium- and high-energy regions of the spectra. 

In Fig.~\ref{fig:backScatterTimes} we present the SP solutions for the ionization [panels (a) and (c)] and rescattering times [panels (b) and (d)] obtained using the SFA and CQSFA theories as indicated in the panels. All solutions exhibit arch-like structures associated with pairs of orbits that coalesce at specific energies. These energies correspond to maxima of the electron's classical kinetic energy, and give kinematic constraints that may be associated with a cutoff in the photoelectron spectra. The highest cutoff energy, at $10U_p$, is associated with the shortest pair of orbits \cite{Becker2015}. The SFA classification of the solutions is presented in Fig.~\ref{fig:backScatterTimes}(a). More precisely, the values of the multiindex $(\beta, m)$ are indicated above panel (a), while the values of index $\alpha$ are indicated within panels (a) and (b). Nonetheless, at first glance, there are a few differences between the SFA and CQSFA results. For example, the ionization and rescattering times calculated using the CQSFA theory are shifted with respect to those obtained using the SFA. Also, the arch-like structures, calculated using the CQSFA, are closed for the low-energy values.

\begin{figure}[h]
    \centering
    \includegraphics[width=1.0\linewidth]{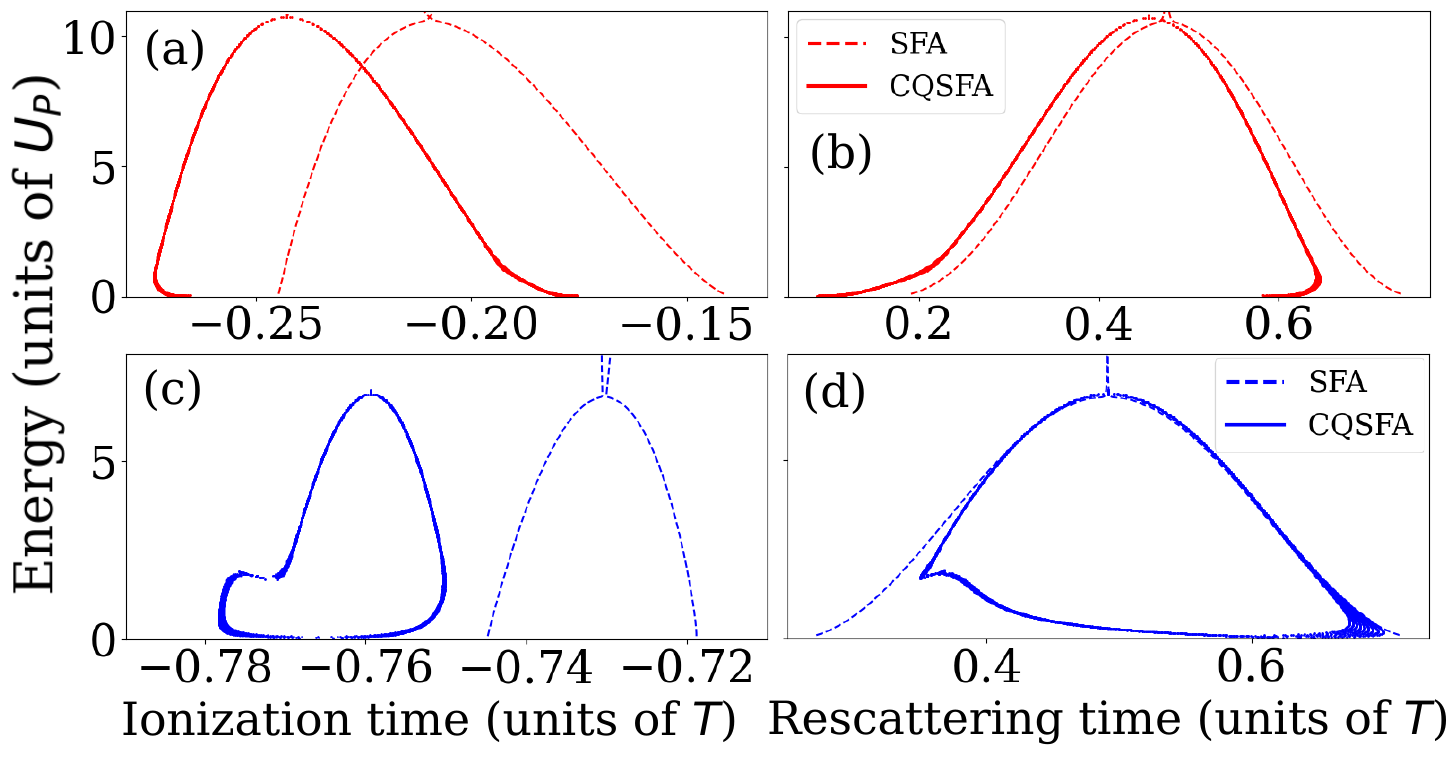}
    \caption{Ionization [panels (a) and (c)] and rescattering [panels (b) and (d)] times for the shortest [panels (a) and (b)] and the second-shortest [panels (c) and (d)] SP solutions calculated using the SFA (dashed lines) and  CQSFA (solid lines) theories.}
    \label{fig:backScatterTimesSpecific}
\end{figure}

In order to better assess the aforementioned differences, in Fig.~\ref{fig:backScatterTimesSpecific} we present the results for the ionization and rescattering times for the shortest [$(\beta, m)=(-1,0)$] [panels (a) and (b)] and  second-shortest [$(\beta, m)=(1,1)$] [panels (c) and (d)] solutions of the SP equations calculated using the SFA (dashed lines) and  CQSFA (solid lines) theories.
Both, the  ionization and rescattering times are shifted when the  Coulomb effects are taken into consideration. For  the  ionization time, this shift is approximately the same for both pairs of solutions, while for the rescattering time, the shift is different for the two pairs. In addition, the effect of the Coulomb potential is generally different for the two solutions  of one pair. The difference between the rescattering times of the solutions of the shortest pair [see Fig.~\ref{fig:backScatterTimesSpecific}(b)] is more pronounced than the corresponding difference for the solutions of the second-shortest pair [see Fig.~\ref{fig:backScatterTimesSpecific}(d)], particularly in the medium- and high-energy regions of the spectrum. For the first return, the difference in travel time  calculated using the SFA and CQSFA theories is small for all values  of the  photoelectron energy. We have checked that this shift becomes more pronounced for later rescattering occurrences. This happens due to the fact that, for the later rescattering events, the electron spends more time close to the core, thus increasing the significance of the Coulomb potential.

\begin{figure}[h]
    \centering
    \includegraphics[width=1.0\linewidth]{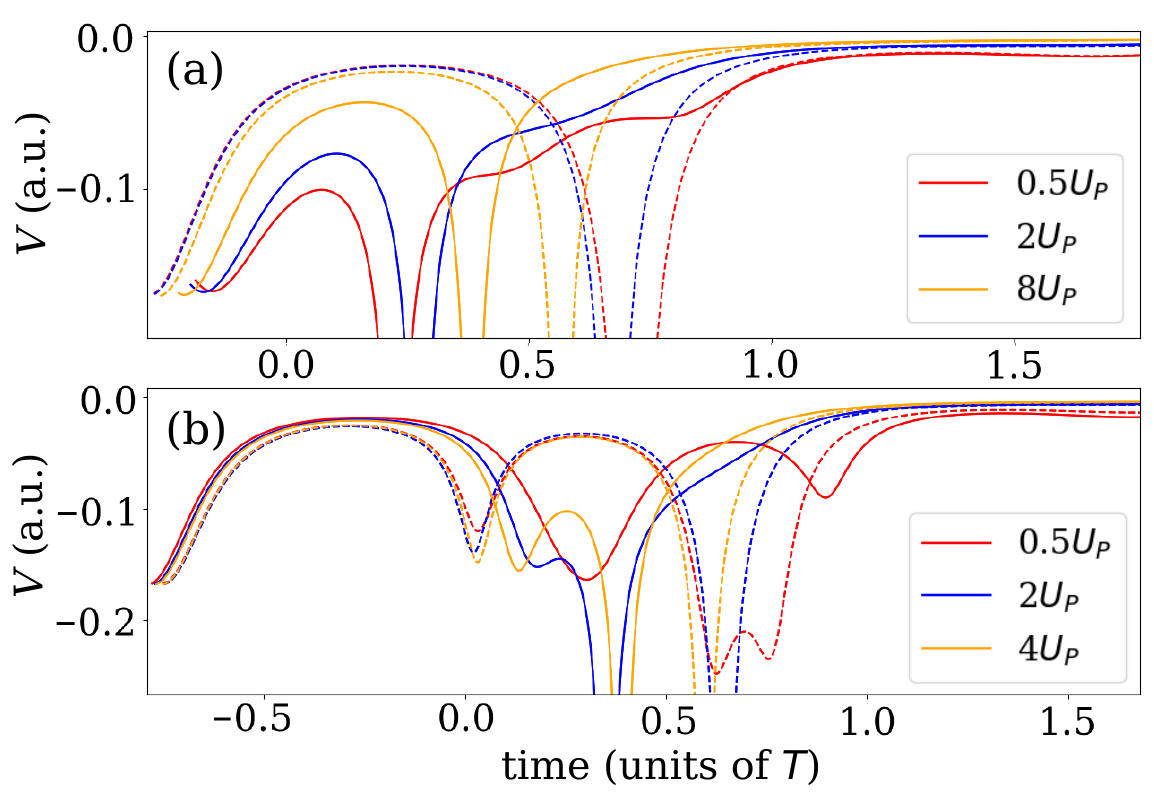}
    \caption{Coulomb  potential as a function of time for the shortest [panel (a)] and the second-shortest [panel (b)] pair  of solutions from Fig.~\ref{fig:backScatterTimesSpecific}. The values of the photoelectron energy are indicated in the panels, while the solid and dashed lines correspond to the  short  and long  trajectories, respectively.}
    \label{fig:Vtime}
\end{figure}

In the low-energy region of the spectrum, the shift of the rescattering  time, due to the Coulomb effect is larger. This can  be explained by the fact that the low-energy electrons spend more time in close proximity to the core than the high-energy electrons. The Coulomb potential energy $V(\ver)$ [see Eq.~\eqref{cqsfa23}] of the liberated electron is sensitive to the small oscillations of the electron subjected to the laser field. In order to illustrate this, in Fig.~\ref{fig:Vtime} we present $V(\ver)$ as a function of time for the shortest [panel (a)] and the second shortest [panel (b)] pair  of solutions. We present the potential for the values of the photoelectron energy $E_{\vp_f}$ as indicated in the legends and for both the short (solid lines) and long (dashed lines) solutions. 
The rescattering events happen around $0.5T$. By comparing the values of the potential during the rescattering  event for different values of the photoelectron energy, we see that, for the high and medium values of the energy, the potential rapidly goes to a large negative  value (see the blue (dark gray) and orange (light gray) lines in Fig.~\ref{fig:Vtime}). On the other hand, for the small value of the photoelectron energy, the  behavior of the potential  during rescattering is  different for the shortest and second-shortest pairs of solutions. In the case of the second-shortest pair, the photoelectron does not closely approach the  core and the potential well is approximately as deep as it is at the tunnel exit [see the red (gray) lines in Fig.~\ref{fig:Vtime}(b)]. In addition, the potential exhibits oscillations which continue long after rescattering (see the red (gray) lines in Fig.~\ref{fig:Vtime}). This explains the fact that the deviation of the rescattering time calculated using the CQSFA from the time calculated using the SFA theory is more  pronounced in the low-energy region.

Finally, the two branches of the second-shortest (as well as for the fourth-shortest) set of solutions approach each other for low-energy in a similar manner as they do for the high-energy cutoff [see Fig.~\ref{fig:backScatterTimesSpecific}(d)]. This can be related to the results shown in Fig.~\ref{fig:momMap}(c). As we already mentioned in Sec.~\ref{sec:mapping}, the sets of saddle point solutions corresponding to the second- and fourth-shortest pairs do not intersect with the $p_{f,y}$ axis. Therefore, as the energy of solutions goes to zero in the direction of emission $\theta_e=0^\circ$, they approach the caustic and coalesce. This low-energy caustic also passes through the origin in the SFA theory and, equivalently, a coalescence of saddles occurs. However, we do not observe the same low-energy coalescence of the rescattering and ionization times since the coalescence is only visible on extremely low energy scales. This happens because the presence of a long-range Coulomb potential implies that the photoelectron requires a certain minimum energy to escape this interaction. 

It is also worth noting that, although the rescattering and ionization times of the third-shortest pair of solutions begin to approach for low energies, they do not actually coalesce. The coalescence occurs, as is the case for the shortest pair, in the half plane with opposite sign of $p_{f,x}$. This is the momentum region occupied by the solutions analogous to the SFA forward scattered solutions. It has been excluded by the restrictions imposed in the CQSFA to match its orbits with the backscattered SFA solutions. For the shortest pair, this region will give rise to the $\mu =0$ solution, and for the third-shortest pair it corresponds to the $\mu =1 $ forward-scattered solutions. There are no forward-scattering orbits corresponding to the remaining pairs. The beginning of this coalescence is visible for the third-shortest pair of solutions because the intersection of the set of saddle point solutions for the third-shortest pair intersects with the $p_{f,y}$ axis at a much smaller perpendicular momentum than the set of saddle-point solutions for the shortest pair.

\begin{figure*}[!htbp]  
    \centering
    \includegraphics[width=0.95\linewidth]{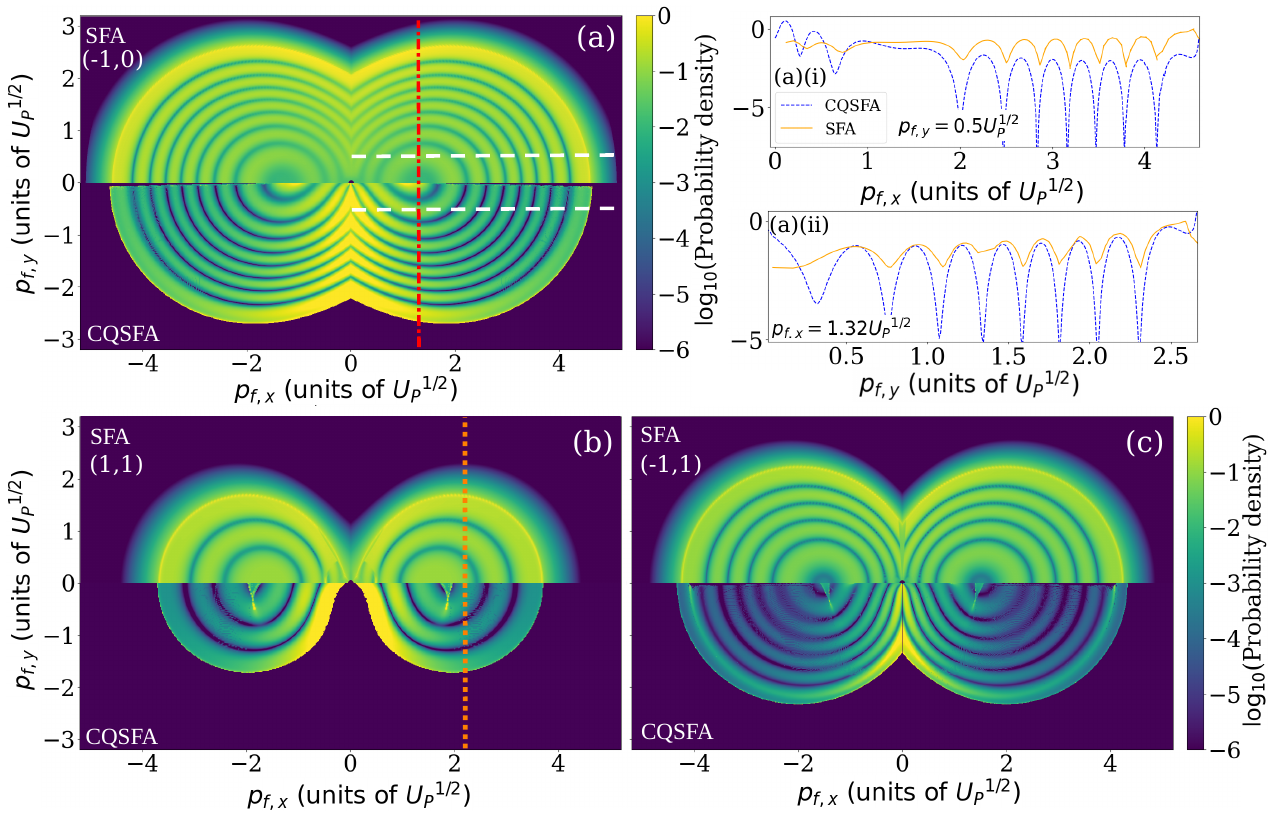}
    \caption{Photoelectron momentum distributions for the shortest [(panel (a)], the second shortest [(panel (b)] and the third shortest [(panel (c)] backward-scattering solutions of the SP equations. The upper-half of the panels corresponds to the SFA-obtained distributions, while the  lower-half of the panels are calculated using the CQSFA theory. Multiindices $(\beta, m)$ which correspond to the SFA-obtained results are indicated in the top-left corner of the panels. Panels (a)(i) and (a)(ii): The linear photoelectron spectra calculated using the SFA (solid orange line) and the CQSFA (dashed blue line) for the values of final momentum denoted by the white dashed and red dash-dotted lines in panel (a). The vertical axis near the color bar refers to the right-hand side as well. The driving-field parameters are the same as in Fig.~\ref{fig:momMap}.}
    \label{fig:backScatterPMDs}
\end{figure*}

\subsection{Photoelectron momentum distributions}

After analyzing the differences between the solutions of the SP equations obtained using the two methods, we now turn our attention to the  differences in the photoelectron momentum distributions. We devote particular attention to the three pairs of solutions with the shortest travel time.

In Fig.~\ref{fig:backScatterPMDs} we present the photoelectron momentum distributions (PMDs) for the shortest [(panel (a)], the  second shortest [(panel (b)] and the third shortest [(panel (c)] backward-scattering solutions of the SP equations. The upper-half of the panels corresponds to the SFA distributions, while the lower-half of the panels are calculated using the CQSFA theory. The results shown in different panels are normalized independently. The solutions obtained using the SFA appear in pairs and the divergent contribution beyond the cutoff has been discarded~\cite{milosevic2002}. The rescattering ridges which appear in the PMDs, calculated using our two theories, appear at roughly the same place in the momentum plane. Each ridge is associated with the maximal classical energy available for a pair of backscattered orbits, and the fringes characterize the interference between the long and short orbits in a pair. They are well-known in the SFA \cite{Becker2015} and have recently been identified in the CQSFA context \cite{rodriguez2023}. Throughout, we see that the SFA fringes are perfectly circular, while their CQSFA counterparts exhibit distortions close to $(p_{f,x},p_{f,y})=(0,0)$. There are also mismatches in the positions of the fringes, which depend on the pair of orbits taken into consideration.  

For the shortest pair of solutions [see Fig.~\ref{fig:backScatterPMDs}(a)], the agreement between the results obtained using the two earlier mentioned theories is good except for the regions $p_{f,x}\in [\pm0.5U_p^{1/2},\pm1.8U_p^{1/2}]$, $|p_{f,y}|< 0.7U_p^{1/2}$. More specifically, the positions of the minima, caused by the interference of the two contributions of the pair appear for approximately the same energy in both cases. In contrast, in the low-energy region, the CQSFA fringes  become tear-shaped and there are mismatches with regard to the SFA. 
The main difference appears for the emission in the direction close to $\theta_e=\pm90^\circ$. In this case, the photoelectron yield calculated using the CQSFA theory is increased with respect to the yield obtained for other values of the emission angle. This increase in the photoelectron yield was discussed in \cite{Maxwell2018} and it is not  present in the photoelectron momentum distribution obtained using the SFA theory. Finally, for energy after the cutoff, the applied SFA theory leads to the probability density which decreases exponentially as a function of the photoelectron energy. On the other hand, for the CQSFA, there are no trajectories with energy beyond the cutoff. In order to include the solutions beyond the classical cutoff, complex trajectories are required. These trajectories can be included as shown in \cite{Maxwell2018b}. 

In order to discuss the agreement between the SFA and CQSFA theories as a function of the photoelectron energy, in Fig.~\ref{fig:backScatterPMDs}(a)(i) we present the photoelectron momentum spectra for a fixed value of the momentum component $p_{f,y}=\pm0.5U_p^{1/2}$ as a function of  the momentum component $p_{f,x}$ [along the white dashed lines in Fig.~\ref{fig:backScatterPMDs}(a)]. Clearly, the agreement between the results obtained using our theories is excellent in the medium- and high-energy parts of the spectrum. As the momentum component $p_{f,x}$ decreases, the agreement between the positions of the minima and maxima calculated using our theories becomes slightly worse. On the other hand, the agreement is not good for $p_{f,x}<U_p^{1/2}$, i.e., in the low-energy part of the spectra. Similar results are presented in Fig.~\ref{fig:backScatterPMDs}(a)(ii) but for a fixed $p_{f,x}=1.32U_p^{1/2}$ [along the red dash-dotted line in Fig.~\ref{fig:backScatterPMDs}(a)] and for changing $p_{f,y}$. In this case, the photoelectron spectrum is shorter, but the discrepancies between the results calculated using our two theories are still not pronounced except for values $p_{f,y}<0.5U_p^{1/2}$. The agreement between the results becomes better with the increase of $p_{f,y}$. Also, the number of predicted interference minima (and maxima) is the same for both theories.

For the second shortest pair of solutions [see Fig.~\ref{fig:backScatterPMDs}(b)], the agreement between the results obtained using SFA and CQSFA theories is much weaker than for the shortest pair. In particular, the positions of the interference minima do not coincide very well indicating that the accumulated phase differences are not the same. Intuitively, this makes sense because the contribution of this pair of solutions is mostly significant in the low- and medium-energy parts of the spectra where the Coulomb effects are particularly pronounced. Furthermore, for $p_{f,x}\approx \pm1.85U_p^{1/2}$ and a small $p_{f,y}$, caustics appear in the results obtained using the CQSFA theory. These caustics are associated with orbits clustering, but are overestimated due to an artefact of the theory, which fails to work well in the vicinity of coalescing saddle points. Similarly as for the shortest pair of solutions, the increase of the  photoelectron yield is predicted by the CQSFA theory, for the emission in the direction close to  the $\theta_e=\pm90^\circ$. In the end, we mention that the solutions of the second-shortest pair (and all other backward-scattering solutions with $\beta=1$) have irregular behavior for the energy close to zero in the SFA approach \cite{milosevic2016a}.   

Finally, for the third-shortest pair, the agreement between the results obtained using our  two theories is reasonable - not as good as for the shortest pair, but better than that observed for the second shortest pair. However, the fringes' positions are mismatched and the CQSFA exhibits a larger number of interference minima (6 against the 5 minima encountered for the SFA). There are also caustics which now appear for $p_{f,x}\approx \pm1.5U_p^{1/2}$ and small $p_{f,y}$. Besides that, the increase of the photoelectron yield in the close to orthogonal direction exists, but it is not pronounced as for the second-shortest pairs. 

\begin{figure}[h]
    \centering
    \includegraphics[width=1.0\linewidth]{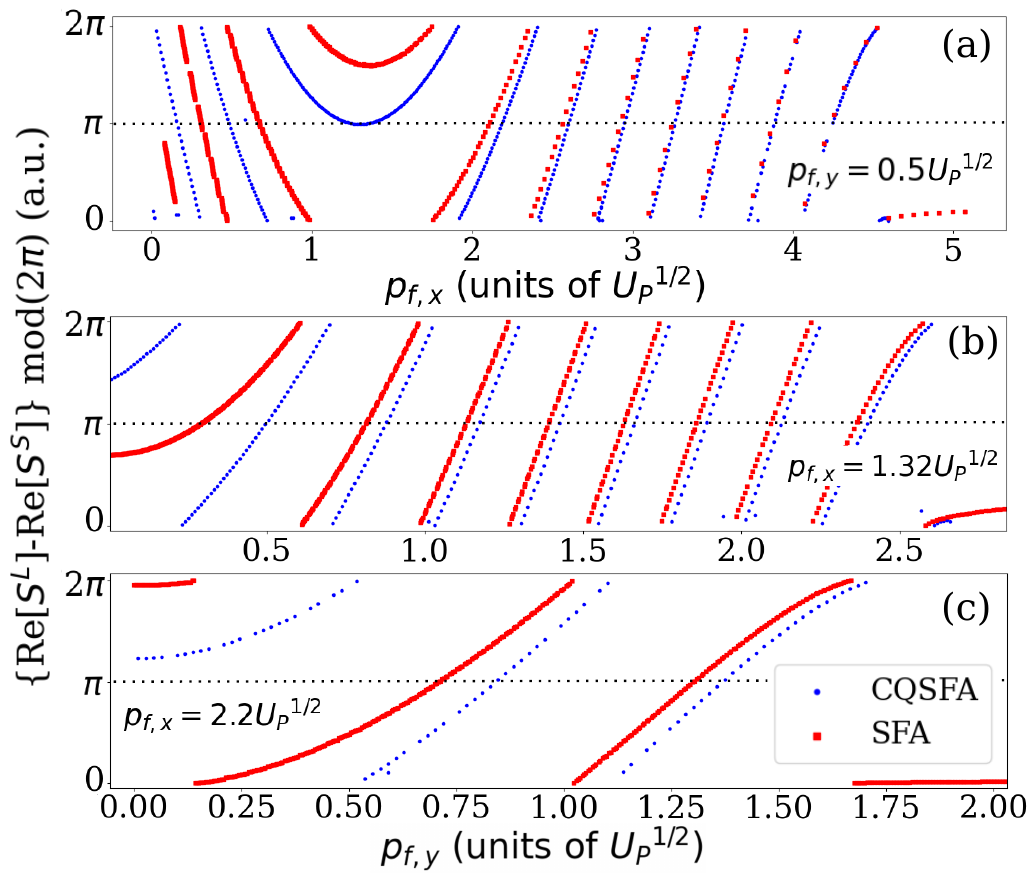}
    \caption{Real part of the difference between the actions $S^L$ and $S^S$, corresponding to the long (superscript '$L$') and short (superscript '$S$') orbits of the shortest [panel (a) and (b)] and second-shortest [panel (c)] pair of solutions, modulus $2\pi$ as a function of the momentum component. The results are calculated using the SFA (dotted red lines) and CQSFA (dotted blue lines) theories along the lines $p_{f,y}=0.5U_p^{1/2}$ and $p_{f,x}=1.32U_p^{1/2}$ for the shortest and $p_{f,x}=2.2U_p^{1/2}$ for the second-shortest pair.}
    \label{fig:actionbackScatt}
\end{figure}
Up to now, we have discussed the rings which appear in the photoelectron momentum distributions calculated using SFA and CQSFA theories. We have concluded that, for the shortest pair of solutions, there is close agreement between the results in the medium- and high-energy parts of the spectra, while it is the case that for the solutions of the second shortest pair, there is disagreement across the entirety of the spectra. In order to explain this, in Fig.~\ref{fig:actionbackScatt} we present the real part of the difference between the actions $S^L$ and $S^S$ modulus $2\pi$ (remainder after dividing by $2\pi$) which correspond to the long (superscript '$L$') and short (superscript '$S$') orbits of the shortest [panels (a) and (b)] and second-shortest [panel (c)] pair of solutions for the fixed values of one momentum component as indicated in the panels [along the white dashed, red dash-dotted and orange dotted lines in Figs.~\ref{fig:backScatterPMDs}(a) and \ref{fig:backScatterPMDs}(b)]. The results are calculated using the SFA (red squares) and CQSFA (blue dots) theories along the lines $p_{f,x}=1.32U_p^{1/2}$ and $p_{f,y}=0.5U_p^{1/2}$ for the shortest and $p_{f,x}=2.2U_p^{1/2}$ for the second-shortest pair. When the difference between the real parts of the actions is equal to an even (odd) multiple of $\pi$, maxima (minima) appear in the spectrum. For the  shortest pair of solutions, the difference $\operatorname{Re} S^L-\operatorname{Re} S^S$ becomes smaller as the longitudinal component of the photoelectron momentum $p_{f,x}$ increases [see Fig.~\ref{fig:actionbackScatt}(a)]. For small values of $p_{f,x}$, the differences between the results obtained using the SFA and CQSFA theories are pronounced. This is in agreement with the results shown in Fig.~\ref{fig:backScatterPMDs}(a)(i). The increased number of minima in the CQSFA spectrum compared to the SFA spectrum can be understood by analysing the behaviour of the difference $\operatorname{Re} S^L-\operatorname{Re} S^S$ in the vicinity of $p_{f,x}\approx1.3U_p^{1/2}$. In this region, the mentioned difference reaches a deeper minimum for the CQSFA than for the SFA, such that it extends to the value $\pi$ and forms an additional interference minimum. On the other hand, for $p_{f,x}=1.32U_p^{1/2}$, the difference $\operatorname{Re} S^L-\operatorname{Re} S^S$ remains almost constant as a function of $p_{f,y}$ except for a low-energy region where the Coulomb effects are pronounced [see Fig.~\ref{fig:actionbackScatt}(b)]. The chosen vertical slice is in the energy region for which the continuous Coulomb interaction is less  significant and the accumulated phase difference is similar for both theories. Finally, for the second-shortest pair of solutions, the difference $\operatorname{Re} S^L-\operatorname{Re} S^S$ remains quite pronounced for the broad range of values of the photoelectron momentum [see Fig.~\ref{fig:actionbackScatt}(c); in accordance with results presented in  Fig.~\ref{fig:backScatterPMDs}(b)]. The difference $\operatorname{Re} S^L-\operatorname{Re} S^S$ still decreases  with the increase of the photoelectron momentum, but, due to the lower position of the cutoff for this solution, the Coulomb effects remain significant for most parts of the photoelectron momentum  distribution. In addition, the shape of the Coulomb potential $V[\ver(t)]$ may influence the action [see Eq.~\eqref{cqsfaAction}]. In particular, for the second-shortest pair, the Coulomb potential $V(\ver)$ exhibits a rich structure prior to the rescattering (see Fig.~\ref{fig:Vtime}), which is not the case for the shortest solution. This implies that, due to this effect, the difference between the CQSFA-calculated action and the SFA-calculated action may be more significant than for the shortest pair of solutions.

Besides the backward-scattering solutions, which are prevalent in the high-energy part of the spectra, the forward-scattering solutions should also be taken  into consideration. The classification of these solutions is presented in Sec.~\ref{sec:mapping}. 

\begin{figure}[h]
    \centering
    \includegraphics[width=1.0\linewidth]{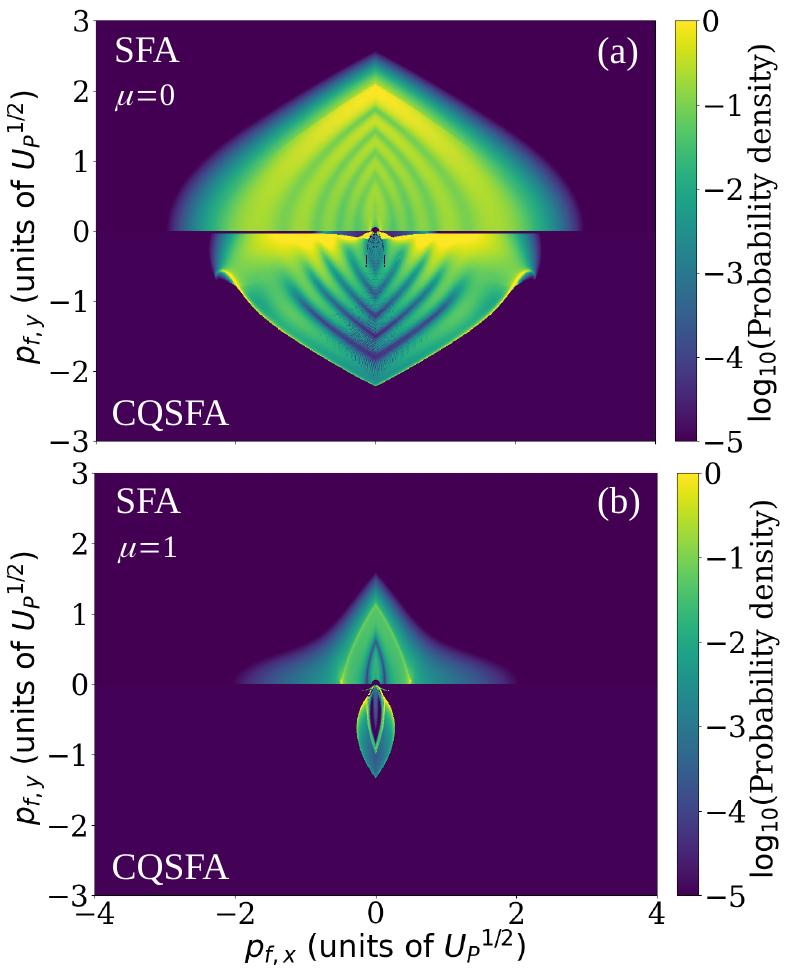}
    \caption{Photoelectron momentum distributions for the shortest [(panel (a)] and the second-shortest [(panel (b)] forward-scattering solutions of the SP equations. The index $\mu$ used in the SFA classification is shown in the top left corner of the panels. The upper-half of the panels corresponds to the SFA-obtained distributions, while the  lower-half panels are calculated using the CQSFA theory. The driving-field parameters are the same as in Fig.~\ref{fig:momMap}.}
    \label{fig:forwScatterPMDs}
\end{figure}
In Fig.~\ref{fig:forwScatterPMDs} we present the photoelectron momentum distributions for the shortest [(panel (a)] and the  second-shortest [(panel (b)] forward-scattering solutions of the SP equations. The upper half of the  panels corresponds to the SFA distributions, while the  lower-half panels  are calculated using the CQSFA theory. The results shown in  different panels are normalized independently. For the shortest pair of solutions, the photoelectron momentum distributions calculated using our theories qualitatively agree well. However, the positions of the  minima caused by the interference of two contributions are not at the same place. This is particularly the case for the minima with a small energy. The fact that the interference minima do not appear at the same place in the photoelectron momentum plane can be explained in a similar manner as for the backward-scattering solutions. In addition, for the emission close to the polarization direction, the CQSFA calculation leads to an increase in the photoelectron  yield in comparison with the one obtained using the SFA theory. Finally, similarly to the backward-scattering  solutions, the CQSFA does not give any nonzero probability  density after the cutoff because  the complex  trajectories are not included. 

In Fig.~\ref{fig:forwScatterPMDs}(b) we see that, for the second-shortest pair of solutions, the differences between the presented results are much more pronounced than for the shortest pair. This is expected due to the fact that the contribution of this solution is significant only in the low-energy part of the spectra. In conclusion, since the contributions of the forward-scattering solutions are significant for low- or medium-energy regions, the Coulomb effects should be taken into consideration.

\subsection{Probing a wider parameter range}

In this section we check if and how the derived conclusions about the SFA and CQSFA results depend on the driving-field parameters, for the first and second shortest pair of orbits. First, we investigate how the difference between the SFA and CQSFA saddle-point solutions for the ionization and rescattering times depends on the laser-field parameters. 
\begin{figure}[h]
    \centering
    \includegraphics[width=1.0\linewidth]{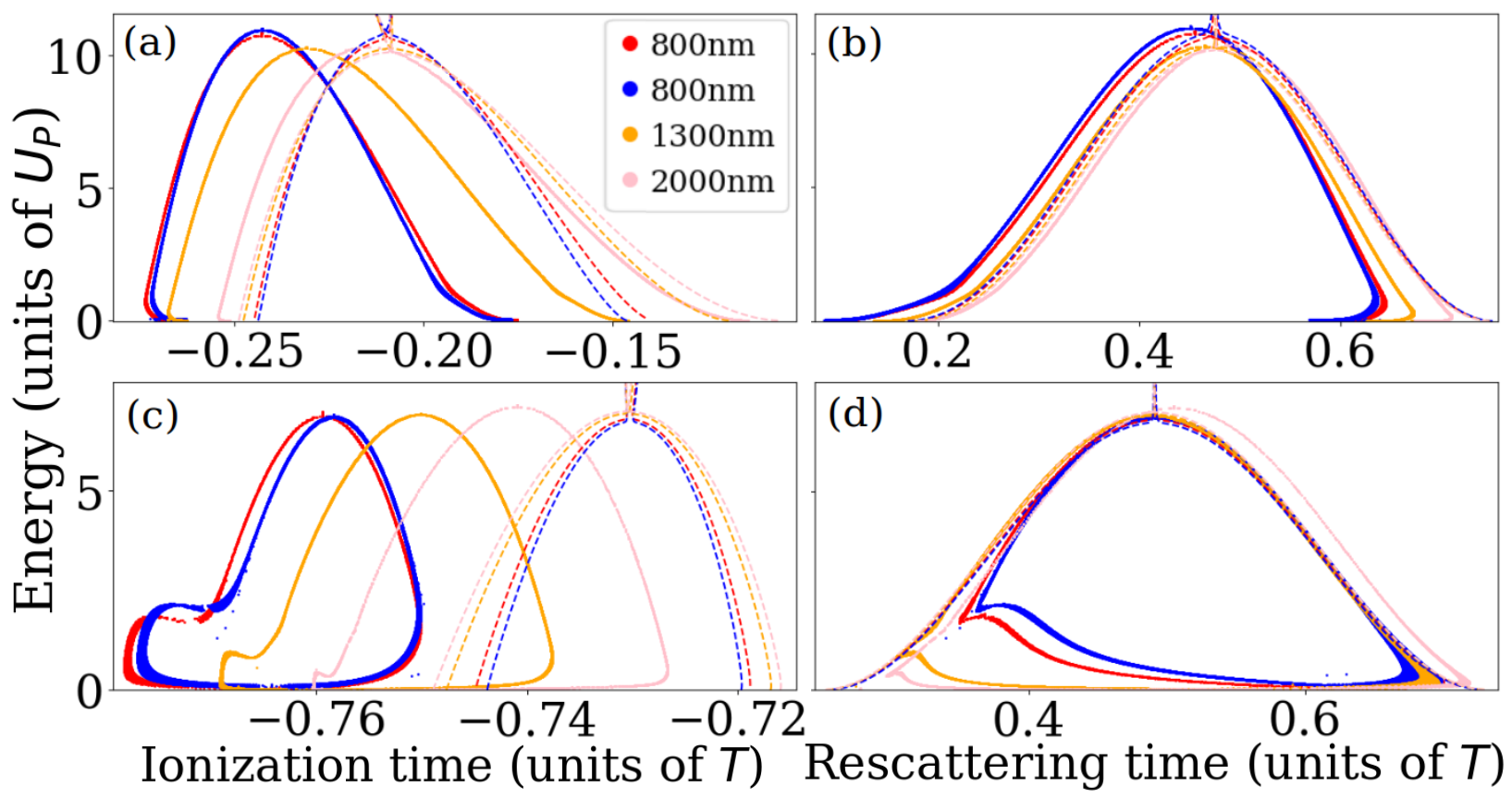}
    \caption{Ionization [panels  (a) and (c)] and  rescattering [panels (b) and (d)] times for the driving-field intensity $I=2\times 10^{14}$~W/cm$^2$ and wavelengths of 2000~nm [pink (bright gray) lines], 1300~nm [orange (light gray) lines], and 800~nm [red (gray) lines], and for the driving field with intensity $I=1.5\times 10^{14}$~W/cm$^2$ and wavelength of 800~nm [blue (dark gray) lines]. Panels (a) and (b) [Panels (c) and (d)] correspond to the shortest (second shortest) pair of the SP solutions. The dashed (solid) lines correspond to the SFA (CQSFA) theory.}
    \label{fig:1new}
\end{figure}
\begin{figure*}[!htbp]  
    \centering
    \includegraphics[width=0.95\linewidth]{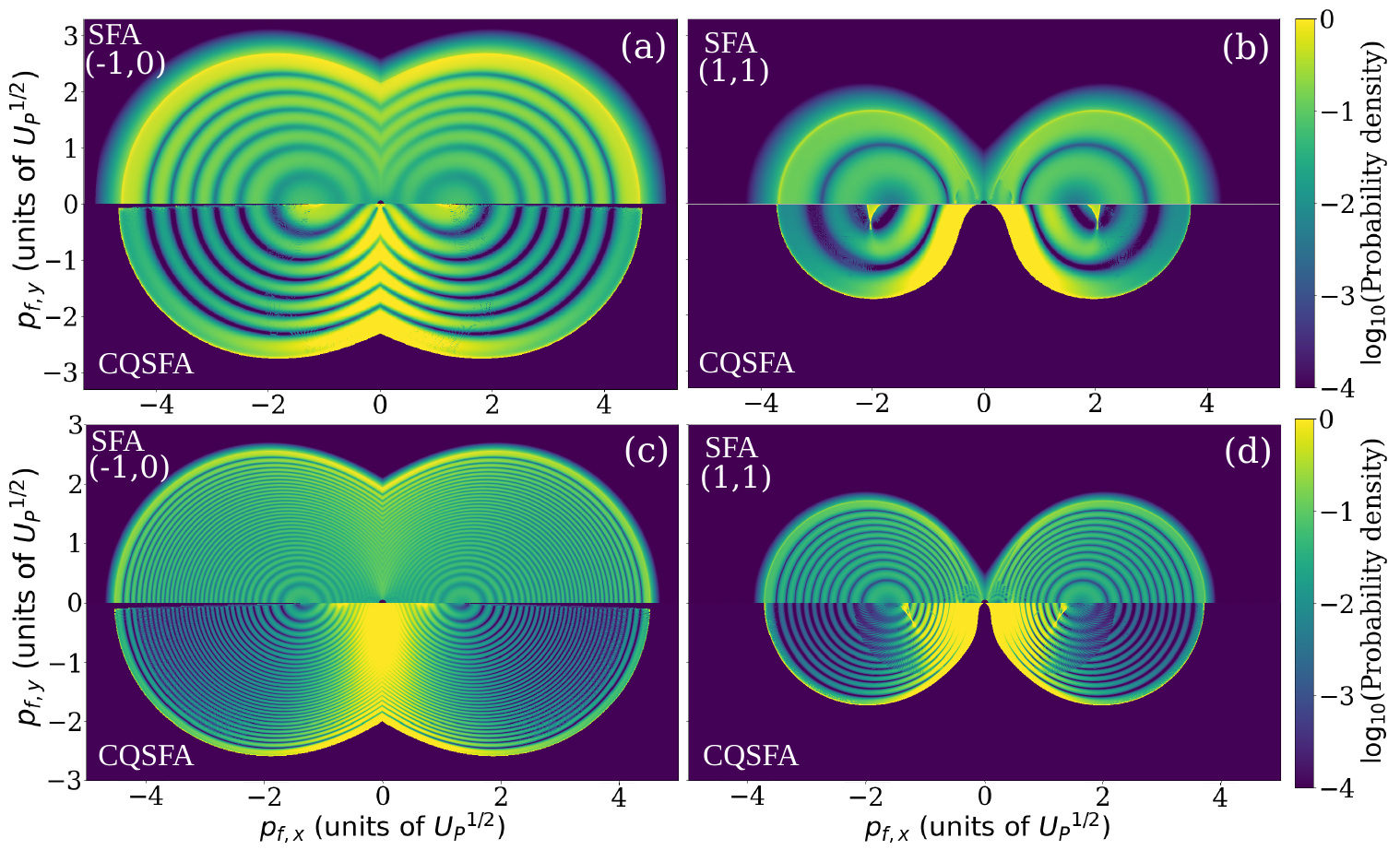}
    \caption{Photoelectron momentum distributions for the shortest [panels (a) and (c)] and the second-shortest [panels (b) and (d)] solutions of the SP equations calculated using the SFA (upper-half of the panels) and the CQSFA (lower-half of the panels) theories. The  driving-field intensity and wavelength are $1.5\times 10^{14}$~W/cm$^2$ and $800$~nm [panels (a) and (b)], and $2\times 10^{14}$~W/cm$^2$ and $1300$~nm [panels (c) and (d)].}
    \label{fig:2new}
\end{figure*}
In Fig.~\ref{fig:1new} we present the ionization [panels  (a) and (c)] and  rescattering [panels (b) and (d)] times for the driving-field intensity $I=2\times 10^{14}$~W/cm$^2$ and wavelengths of 2000~nm [pink (bright gray) lines], 1300~nm [orange (light gray) lines], and 800~nm [red (gray) lines]. In addition, the blue (dark gray) lines correspond to the saddle-point solutions obtained using the driving field with intensity $I=1.5\times 10^{14}$~W/cm$^2$ and wavelength of 800~nm. The dashed (solid) lines denote the solutions obtained using the SFA (CQSFA) theory. Figures~\ref{fig:1new}(a) and (b) [Figures~\ref{fig:1new}(c) and (d)] correspond to the shortest (second-shortest) pair of the SP solutions. The change of the driving-field parameters affects the difference between the saddle-point solutions obtained using our two theories which is particularly visible for the ionization time. In particular, as the driving-field wavelength increases, the difference between the results becomes less pronounced. For example, the difference between the ionization-time SP solutions for the wavelength of 800~nm is far larger than for the wavelength of 2000~nm, for the same laser-field intensity [cf. the difference between the red (gray) solid and dashed lines and the pink (bright gray) solid and dashed lines in Figs.~\ref{fig:1new}(a) and (c)]. This is more pronounced for the shortest pair of the SP solutions. Moreover, the difference between the SFA and CQSFA results is less affected by the change in the driving field intensity [cf. the difference between the red (gray) solid and dashed lines and the blue (dark gray) solid and dashed lines in Figs.~\ref{fig:1new}(a) and (c)]. This happens because the electron excursion in the laser field is directly proportional to the square root of the driving-field intensity and directly proportional to the square of the applied-field wavelength. Because the Coulomb effects fade away at the large distance of the core, the agreement between the SFA and CQSFA results becomes better much more rapidly with the increase of the wavelength than with the increase of the intensity.  Nonetheless, there are subtle differences observed for different driving-field intensities, in particular for the second shortest pair of orbits. For the lower intensity, the range of ionization times is slightly narrower, and the low-energy ends of the arch-like structures associated with the rescattering times close at marginally larger values. This is due to the increased influence of the Coulomb potential. 
Finally, we note that the curves which correspond to the ionization and rescattering times as functions of the photoelectron energy are closed regardless of the values of the driving-field parameters for the second-shortest pair of the SP solutions. As the photoelectron energy approaches zero, the two solutions coalesce. This coalescence is not sudden since the photoelectron requires a certain minimum energy to escape the Coulomb interaction. The reason for this behaviour being restricted to the second-shortest pair (it also occurs for the fourth-shortest pair which is not shown in Fig.~\ref{fig:1new}) was elaborated on earlier using the results shown in Fig.~\ref{fig:momMap}(c).

Let us now investigate how the difference between the photoelectron momentum distributions obtained using the SFA and CQSFA theories is affected by the change of the applied-field parameters. In Fig.~\ref{fig:2new} we present the photoelectron momentum distributions for the shortest [panels (a) and (c)] and the second-shortest [panels (b) and (d)] solutions of the SP equations calculated using the SFA (upper-half of the panels) and the CQSFA (lower-half of the panels) theories. The driving-field intensity and wavelength are $1.5\times 10^{14}$~W/cm$^2$ and $800$~nm [panels (a) and (b)], and $2\times 10^{14}$~W/cm$^2$ and $1300$~nm [panels (c) and (d)], respectively. For both cases, the agreement between the calculated results is good, and the differences are mainly restricted to narrow regions in the photoelectron momentum plane. However, the agreement is better for the larger intensity and longer wavelength [cf.  panel (a) with panel (c), and panel (b) with panel (d)]. For the case of the long wavelength and the high intensity, the minima caused by the interference of the contributions of the two solutions of one pair coincide very well for both the shortest and second-shortest pair. The only difference appears in the narrow region around $(p_{f,x},p_{f,y})=(\pm1.4U_p^{1/2},0)$ for the shortest and around $(p_{f,x},p_{f,y})=(\pm1.6U_p^{1/2},0)$ for the second-shortest pair. On the other hand, for the smaller intensity and shorter wavelength, the agreement is not excellent, particularly for the second-shortest pair of the SP solutions in which case the minima obtained using the SFA and CQSFA theories do not appear at the same place in the photoelectron momentum plane. Finally, we mention that we did not choose the laser field with the wavelength of 2000~nm for our illustration since in this case, the number of interference minima is so large that it is difficult to access the agreement between the results obtained using our theories.

\section{Conclusions}\label{sec:conclusion}

In this work, we compare two semianalytical theories based on the saddle-point method, which allows one to get a clear insight into the underlying physics of the strong-field ionization process. The first one is the strong-field approximation (SFA) which assumes that the driving field is so strong that the Coulomb attraction between the liberated electron and the residual ion can be neglected during the electron propagation in the continuum. The second theory is the Coulomb quantum-orbit strong-field approximation (CQSFA) for which the Coulomb potential and the driving-field potential are treated equally. These theories are structurally very different. The former is a Born-type-expansion theory for which the rescattering times and conditions are well-defined because the interactions with the parent ion during rescattering are localized at a single point, namely the origin. On the other hand, the latter has a Coulomb distorted continuum which consequently makes it impossible to clearly define when the rescattering starts or ends.  
 
To compare the results obtained using the SFA and CQSFA is not as trivial a task as it sounds at first. In particular, for a systematic comparison, it is not sufficient to remove the Coulomb coupling or make it short-range due to the structural differences between the two theories. These differences make it hard to establish a one-to-one correspondence between the orbits obtained with each method. In order to make this comparison, we have identified the CQSFA trajectories that mimic the behavior of the Born-type SFA trajectories with a single act of rescattering. This means that among all orbits that appear in the  CQSFA theory, we have extracted only those with SFA counterparts. We also provide an example of an orbit for which this correspondence does not hold, although CQSFA orbits with no SFA counterparts have been studied in far more detail in previous publications \cite{Lai2017,Maxwell2017,Maxwell2018,rodriguez2023}. 

Here, we have also focused on the simplest possible system: hydrogen in a linearly polarized monochromatic field. This was purposefully done not to introduce other effects, which would mask what we intended to single out. Focal averaging may distort the interference patterns of interest \cite{Maxwell2016,Maxwell2022}, or be a source of incoherence \cite{Maxwell2021a}. Short pulses will exhibit unequal cycles, which would give rise to rescattering ridges of unequal energies \cite{Spanner2004,Borbely2019} and holographic patterns that may look quite different from those obtained with longer pulses or monochromatic fields  \cite{Murakami2020,Yuan2021,Taoutioui2022}. Averaging over the carrier-envelope phase will also weaken or wash out subtler interference patterns \cite{Bergues2011,Moeller2014}. Bi- or polychromatic fields would break symmetries and lead to additional solutions to those in Table \ref{tab:orbits}, even for very weak fields \cite{Rook2022}.  These distortions increase for two-color fields with comparable amplitudes \cite{Arbó2015}. For a thorough study of symmetries in the context of the SFA, see \cite{Habibovic2021}. Furthermore, although the CQSFA has been successfully used to study rare gases or small molecules \cite{Maxwell2020,Kang2020,Werby2021,Werby2022} employing effective potentials, the extra phase shifts arising from them are not useful to our problem. They have been used, however, to detect the parity of orbitals \cite{Kang2020}. Different targets will also introduce additional momentum biases stemming from the bound-state geometries \cite{Krecinic2018,Borbely2019,Bray2021}.

In the present  paper, we have found that the agreement between the results obtained using our theories is fairly good if the solutions used in the CQSFA theory are restricted to only those which we expect to have a counterpart in the SFA theory. This happens because the compared orbits accumulate similar phases in the continuum, although their dynamics can be very different. For example, before the rescattering, the SFA trajectories are one-dimensional for a linearly polarized driving field, while for the CQSFA theory, the corresponding orbits are always two-dimensional. More generally, the SFA trajectories follow the field, which is not the case with their CQSFA counterparts due to the always present Coulomb interaction.

In order to systematically investigate the influence of the Coulomb effects, we have analyzed the photoelectron momentum distributions for different solutions of the saddle-point equations. We have found that the positions of the ridges for different solutions are approximately the same for both theories, while the positions of the minima caused by the interference of the contributions may be different, depending on the value of the photoelectron energy. Whether the positions of these minima are reproduced well in the SFA or not depends on two counterbalancing effects: the time the electron spends in the continuum and the electron's energy. A longer time in the continuum means that the accumulated  Coulomb phase will be larger as well, but a higher kinetic energy means that the electron will be less sensitive to the Coulomb potential. These effects are clearly visible in the photoelectron momentum distributions for the three shortest pairs of orbits. The best agreement is for the shortest pair because the corresponding trajectories have a short travel time in the continuum and the corresponding energy is the highest. The worst agreement is with the second shortest pair, because the travel time is longer than for the shortest pair, and the corresponding energy is the lowest. Finally, the third shortest pair has a reasonable agreement, because, although the orbits spend a longer time in the continuum, their energy is higher than those of the second-shortest pair, which partly counterbalances this effect. These conclusions have been confirmed by explicit calculation and comparison between the accumulated phases in the SFA and CQSFA theories.

In conclusion, even though the SFA theory does not include many possible cases of the electron trajectories and the included trajectories are treated by neglecting the existing Coulomb potential, it is still a reasonably good approximation which accounts for a lot of the key dynamics. The obtained results are usually sufficiently accurate for various purposes. Finally, one expects the level of agreement between the theories to depend on the parameters of the driving laser field and the potential. A systematic study of this dependence could help precisely establish the range of values in the parameter space for which the application of the SFA theory is acceptable. Also, the study of the structural differences between the two theories presented in the current paper could be utilized as the groundwork for establishing future hybrid theories.

\acknowledgments
D. H. would like to thank the University College London for its kind hospitality. This work was partly funded by grants No.\ EP/J019143/1 and EP/T517793/1, from the UK Engineering and Physical Sciences Research Council (EPSRC). D. H. and D. B. M. gratefully acknowledge support by the Ministry for Science, Higher Education, and Youth, Canton Sarajevo, Bosnia and Herzegovina.

\end{document}